\documentstyle[prb,aps]{revtex}

\begin{document}
\title{Tunneling driven tilt modes of the O octahedra in La$_{2-x}$Sr$_{x}$CuO$_{4}$%
: strong dependence on doping}
\author{F. Cordero}
\address{CNR, Area di Ricerca di Tor Vergata, Istituto di Acustica ``O.M. Corbino``,\\
Via del Fosso del Cavaliere 100, I-00133 Roma, and INFM, Italy}
\author{R. Cantelli}
\address{Universit\`{a} di Roma ``La Sapienza``, Dipartimento di Fisica, P.le A.\\
Moro 2, I-00185 Roma, and INFM, Italy}
\author{M. Ferretti}
\address{Universit\`{a} di Genova, Dipartimento di Chimica e Chimica Fisica,\\
Via Dodecanneso 31, I-16146 Genova, and INFM, Italy}
\maketitle

\begin{abstract}
The anelastic spectrum of La$_{2-x}$Sr$_{x}$CuO$_{4}$ ($x=$ 0, 0.008, 0.019,
0.032) has been measured down to $1.5~$K, in order to see the effect of
doping on the intrinsic lattice fluctuations already found in stoichiometric
La$_{2}$CuO$_{4}$, and identified with tunneling driven tilt modes of the O
octahedra. Slight doping with Sr causes a drastic increase of the transition
rates and relaxation strength of the tunneling systems. The influence of
doping on the relaxation rate is interpreted in terms of direct coupling
between between the tilts of the octahedra and the hole excitations.
However, the observed fast dependence of the rate on temperature cannot be
explained in terms of the ususal models of coupling between a tunneling
system and the conduction electrons.
\end{abstract}

\twocolumn

\section{Introduction}

Among the several noticeable properties of the superconducting cuprates, the
interplay between the lattice and the electrons is particularly intriguing
and gives rise to a variety of phenomena which are difficult to be
systematized. On one side, there are several indications of separation of
the charge carriers into nanodomains which are superconducting and
antiferromagnetic,\cite{SM94} and the charge phase separation is thought to
be an essential ingredient of the mechanisms leading to high $T_{c}$
superconductivity.\cite{EKZ97,SCD98,Phi97} There is also growing evidence
that the superconducting and antiferromagnetic domains consist of stripes
with a periodicity which is incommensurate with the lattice and dependent on
doping, suggesting that these phenomena exclusively arise from interactions
between the charge carriers.\cite{EKZ97} Indeed, although a lattice
modulation should correspond to the charge modulation, superlattice Bragg
peaks are hardly observed in the superconducting cuprates, unless the
modulation is pinned at a commensurate wave vector by sufficient disorder in
the ion sizes in the La sublattice.\cite{TAI} The lack of definite
superlattice peaks is attributed to the fluctuating nature of these stripes
and their short coherence length.

On the other side, there is growing evidence that the lattice of
perovskite-type materials, including the layered superconducting cuprates,
locally has a lower symmetry than that evinced from diffraction experiments.
Such anomalous lattice fluctuations can be reproduced in models of highly
anharmonic lattice dynamics, e.g. due to the high polarizability of the O
ions,\cite{Bus99} without explicit introduction of mobile charge carriers.
Local lattice fluctuations, charge separation or charge density waves
instabilities and the mechanisms of electron-phonon interaction may well be
connected to each other, but no clear relationship between these phenomena
has been established yet.

Previous anelastic spectroscopy\cite{61} and joint anelastic and NQR
measurements\cite{68} showed that the tilt modes of the O octahedra in
undoped La$_{2}$CuO$_{4}$ present pseudodiffusive dynamics, namely
collective thermally activated hopping between the minima of a multiwell
potential. In what follows we are concerned with the faster motion whose
effects in the anelastic spectra are observed at liquid He temperature, and
imply a tunneling driven motion\cite{61} of more local character. The
present results show a drastic increase of the characteristic frequencies of
these pseudodiffusive modes in La$_{2-x}$Sr$_{x}$CuO$_{4}$ with slight
doping, indicating that these tilt modes are directly coupled with the
electronic excitations.

\section{Experimental}

We measured the complex Young's modulus of La$_{2-x}$Sr$_{x}$CuO$_{4+\delta }
$ at small doping down to $1.2~$K. The samples, with nominal compositions $%
x=0$, 0.0075, 0.015 and 0.03, were prepared by standard solid-state reaction
from powders first treated for 18~h in air at 1050~$^{\text{o}}$C, checked
by X-ray diffraction, pressed, sintered for 18~h in air at 1050~$^{\text{o}}$%
C as described in Ref. \onlinecite{DFF94} and cut as bars approximately $%
40\times 4\times 0.6$\ mm$^{3}$. The Sr content and homogeneity of the
samples where checked from the transition between the tetragonal (HTT) and
the orthorhombic (LTO) structure, which occurs at a temperature $T_{t}$
linearly decreasing with doping\cite{Joh97} as $T_{t}\left( x\right) =\left(
535-2180\,x\right) ~$K. The effect of the transition on the complex Young's
modulus was measured after outgassing from interstitial O, which also lowers
the transformation temperature. The transition from HTT to LTO produces a
huge softening of the in-plane shear modulus, whose temperature dependence
in measurements on single crystals could be fitted within the HTT phase by
using an appropriate Landau free energy.\cite{SMM} However, unlike a usual
displacive transformation, the softening proceeds also in the LTO phase,\cite
{LLN} instead of recovering. For this reason, the step in the Young's
modulus is rather broad and it is not straightforward to determine an exact
transition temperture $T_{t}$. Concomitantly with the modulus drop, the
absorption increases, so that it is possible to determine the occurrence of
the transition also from the absorption step. We assumed that in the Sr-free
sample it is $T_{t}=535~$K (Ref. \onlinecite{Joh97}), which falls exactly at
half of the modulus step, and determined $T_{t}$ for $x>0$ by shifting the
temperature scale in order to overlap the steps in the modulus (as shown in
Fig. 1) or in the absorption. In this way, for the three Sr-doped samples we
find $T_{t}=518\pm 3~$K, $493.5\pm 1.5~$K and $464\pm 4~$K, corresponding to 
$x=0.008\pm 0.0015$, $0.019\pm 0.001$ and $0.032\pm 0.002$ respectively.

From Fig. 1 it appears that, except for the sample with $x=0.008$, the
transition of the Sr-doped samples is sharper than that of the Sr-free
sample, ensuring that the broadening is intrinsic and Sr is homogeneously
distributed. The broader transition of the sample with $x=0.008$ may
indicate some inhomogeneity in the Sr distribution at such very low doping
level.

The imaginary part (acoustic absorption, or reciprocal of the mechanical $Q$%
)\ and real part (Young's modulus $E$) of the dynamic modulus where measured
by electrostatically exciting the flexural vibrations of the bars suspended
on thin wires in correspondence with the nodal lines. The resulting
anelastic spectra provide information on the low frequency dynamical
processes, mainly of relaxational or diffusive character, i.e. described by
a correlation function of the form $\exp \left( -t/\tau \right) $, where $%
\tau $\ is the relaxation (correlation) time. In the ideal case of
independent units described by the same $\tau $, one has pure Debye
relaxation:\cite{NB} 
\begin{equation}
Q^{-1}=\Delta \frac{\omega \tau }{1+\left( \omega \tau \right) ^{2}}\,,\quad 
\frac{dE}{E}=-\Delta \frac{1}{1+\left( \omega \tau \right) ^{2}}\,,
\label{Debye}
\end{equation}
which are a peak in the absorption and a step in the modulus dispersion
centered at the temperature for which $\omega \tau =1$, where $f=\omega
/2\pi $ is the sample vibration frequency; the relaxation rate $\tau ^{-1}$
is generally an increasing function of temperature, and therefore a
temperature scan of the complex modulus presents the absorption peak and
modulus step at a temperature which decreases with decreasing the measuring
frequency, or with increasing relaxation rate. The relaxation strength $%
\Delta $ is proportional to the fraction of the relaxing units, which the
are dynamically tilting octahedra, for the processes which we are concerned
with. The measurements were made on cooling, but the whole relaxation
spectrum of the Sr-free samples is reproducible below room temperature also
on heating (unpublished results%
); since Sr-free samples exhibit the same absorption peaks of Fig. 1,
although with different intensities or temperatures, we expect stability
against thermal cycling below room temperature also in the present case.

In the as-prepared condition, the presence of interstitial O was clearly
detected in the anelastic spectra of all the samples. The content of excess
O was reduced by heating in high vacuum up to 790$~$K.

\section{Results}

Figure 2 presents the anelastic spectrum of the sample with $x=0.019$ after
O reduction, measured during the same cooling run at three excitation
frequencies corresponding to the 1st, 3rd and 5th flexural modes. The whole
spectrum shifts to higher temperature for higher measurement frequency,
indicating that all the processes below $300~$K are thermally activated with
relaxation rates $\tau ^{-1}\left( T\right) $ which are increasing functions
of temperature. Since the resonant frequency for flexural vibrations is
proportional to $\sqrt{E}$, the relative change of Young's modulus $dE/E$\
is given by $\left[ f\left( T\right) /f_{0}\right] ^{2}-1$; for each mode,
the reference frequency $f_{0}$\ has been chosen in order to let $dE/E$\ to
coincide for all the modes near $30~$K, where the relaxation processes are
negligible.

The peak around $160~$K, labeled T, is due to the cooperative tilt motion of
the octahedra; such a process has been measured in La$_{2}$CuO$_{4}$ also as
a maximum in the $^{139}$La NQR relaxation rate, and has been discussed in
detail in Ref. \onlinecite{68}. It has been interpreted in terms of
propagating tilt waves, with an effective activation energy ($2800~$K for $%
x=0$) which is higher than the barrier separating different minima of the
multiwell local potential felt by each octahedron, due to the cooperative
character of the motion of the octahedra (the calculated barriers are of the
order of few hundreds K, see Ref. \onlinecite{CPK,BMF}). Peak T in Fig. 1 is
similar to that measured in undoped La$_{2}$CuO$_{4}$, but with a smaller
amplitude and a slightly smaller activation energy ($2600~$K).

One of the two minor peaks between 60 and 100$~$K may correspond to a
similar anelastic peak which has been observed at higher Sr and Ba doping,
and assigned to an electronic relaxation;\cite{GKB93} this processes will be
dealt with in a separate work.\cite{CCC00}

We will focus on the relaxation which appears below $10~$K and is labeled A
in Fig. 2, again due to intrinsic lattice fluctuations,\cite{61,nota}
presumably in correspondence to the tilt waves, whose collective motion is
frozen at these temperatures. The central result, shown in Fig. 3, is the
dependence of peak A on doping with Sr, after reducing the content of excess
O. The peak measured exciting the 5th vibration mode (6-9$~$kHz) is centered
at $5~$K for $x=0$, but already at $x=0.008$ the maximum has shifted to $4~$%
K and increased in intensity by 3.5 times; at $x=0.019$ the maximum is
shifted below 1.4$~$K, and at $x=0.03$ the peak is no more visible within
the experimental temperature window. It cannot be said whether the
relaxation magnitude continues increasing for $x>0.008$, since only the tail
of the peak is observable at higher doping levels.

We can state that {\it doping has a very strong effect on the low
temperature relaxation: }it {\it increases its intensity} and shifts it to
lower temperature, which implies through Eq. (\ref{Debye}) an {\it %
acceleration of the pseudodiffusive dynamics of the octahedra}. A
progressive shift of peak A to lower temperature on doping with excess O had
already been observed,\cite{61} but small and obscured by a concomitant
decrease of the peak intensity due to the blocking effect of interstitial O
(see later). The present measurements show that the enhancement of the
relaxation rate with doping is huge and also the relaxation magnitude
increases, when the complications from blocking effects are reduced.

The minor step visible below 7$~$K at $x=0.032$ in Fig. 3, labeled B, is
distinct from the main relaxation process A, since it is present also at $%
x=0.019$, and has a different behavior on O doping. This is shown in Fig. 4,
where the absorption is reported both in the outgassed state (same symbols
of Fig. 3) and in the as-prepared state (thick lines without experimental
points). Interstitial O is present in the samples prepared in air, and its
concentration is estimated as $\delta \simeq 0.005$ in our Sr-free samples.%
\cite{67} We estimate $\delta >0.002$ and $0.001$ for $x=0.019$ and $0.032$,
respectively, based on the intensity of the relaxation process\cite{63} due
to the diffusion of excess O (not shown here).

The above O concentrations are considerably smaller than those of the Sr
dopant, but have a profound effect on the relaxation spectrum, since each
interstitial O atom\ blocks several surrounding octahedra into fixed
orientations, inhibiting their pseudodiffusive motion within the multiwell
potential. Indeed, peak T is completely suppressed in all the samples in the
as-prepared state (not shown for $x\neq 0$; see Ref. \onlinecite{61} for $%
x=0 $). The effect of excess O on peak A is less drastic, indicating that
the latter mechanism involves a smaller number of octahedra than peak T; it
consists of a depression of the intensity and a shift to lower temperature.
The peak shift to lower temperature is due to the holes doped by excess O,%
\cite{61} as discussed later. This is clearly seen in Fig. 4 for $x=0$, and
to a smaller extent for $x=0.019$ below $7~$K; instead, the step labeled B
remains unaffected by the presence of excess O for both $x=0.019$ and $0.032$%
, indicating that its nature is different from that of the major peak. In
the following we will discard the small step-like feature B with respect to
the main peak A.

\section{Discussion}

\subsection{Collective tilt modes of the octahedra}

The low temperature relaxational response of the tilt modes of the octahedra
may be due to the formation of fluctuating LTT domains in the LTO\ structure.%
\cite{61} Such domains may correspond to the locally correlated atomic
displacements found in a model anharmonic lattice of perovskites,\cite{Bus99}
but the case of La$_{2-x}$Sr$_{x}$CuO$_{4}$ and possibly of other layered
cuprates may be different, since they consist of O\ polyhedra which are
connected in two dimensions instead of three. It is even possible that the
bidimensional array of connected octahedra is describable as a
one-dimensional system.\cite{Mar93b} Although recent extended X-ray
absorption fine structure (EXAFS)\cite{HSH96} and atomic pair distribution
function (PDF)\cite{BBK98} measurements of La$_{2-x}$Sr$_{x}$CuO$_{4}$
exclude a prevalent LTT local tilt at small $x$, we think that the model by
Markiewicz\cite{Mar93b} of a LTO\ structure arising from a LTT ground state
(dynamic Jahn-Teller phase) provides a good framework for analyzing the
dynamics of the connected octahedra without charge doping. In the LTT
pattern the octahedra are tilted about axes passing through the in-plane
Cu-O bonds, so that the tilt of an octahedron determines the staggered tilts
of all the octahedra in the same row perpendicular to the rotation axis;
instead, the neighboring rows are weakly coupled, since they share O atoms
which remain in the Cu plane. In this manner, the system becomes a
one-dimensional array of rows of octahedra, with each row described by a
single angle of staggered tilts. The one-dimensional non-linear equation of
motion of the tilt angles admits solitonic solutions corresponding to
propagating LTO walls which separate LTT domains. The density, thickness and
speed of the walls depend on the lattice potential, and an array of closely
spaced walls produces an average LTO lattice.\cite{Mar93b} Although the
mobile charges do not play any role, this picture closely corresponds to the
phase of superconducting LTO-like and antiferromagnetic LTT-like stripes
that is proposed to be common to all the superconducting cuprates;\cite{BSL}
these tilt waves could well provide a preferred locus for the charge
stripes, at least at lower doping.

The combined anelastic and NQR relaxation measurements of the peak T have
been interpreted\cite{68} in terms of the propagation of LTO and LTT
solitonic tilt waves, similar to those proposed by Markiewicz. Solitonic
solutions are generally found in one-dimensional models of anharmonic
lattices;\cite{Aub76,BBB87,BR91} their spectral density contains a component
of pseudodiffusive motion,\cite{Aub76} which produces a central peak
observable in NQR experiments\cite{TR87} and anelastic relaxation\cite{68}
of the form of Eq. (1). These solitonic waves are walls between different
domains,\cite{Aub76,BBB87,BR91} and in the present case of very low doping,
where both diffraction and local probes\cite{HSH96,BBK98} indicate a
prevalence of LTO tilts, the tilt waves should consist of LTT walls
separating LTO domains. They should differ, however, from the usual twin
walls between LTO domains,\ likely responsible for the increased dissipation
below the HTT to LTO transformation.\cite{LLN,61} The possibility should be
explored that tilt waves exist, not necessarily separating different LTO
domains related to each other by a rotation of 90$^{\text{o}}$, but that can
form within a same LTO domain; they would be relatively stable, since the
LTT pattern is a local minimum of the potential energy.\cite{CPK,CW} Another
difference between the tilt waves responsible for peak T and the twin walls
can be that they have little or no correlation along the $c$ axis (making
difficult their observation by diffraction).

\subsection{Peak A and the tunneling-driven tilting of the octahedra}

The mechanism responsible for peak A should involve the combined motion of
few octahedra or even few O atoms, as indicated by the faster dynamics and
the reduced blocking effect of O with respect to peak T (involving whole
rows of octahedra and completely frozen at the temperature of peak A).

As mentioned above, the relaxing units could consist of fluctuating LTT
domains in the LTO\ structure,\cite{61} corresponding to the locally
correlated atomic displacements proposed for the anharmonic lattice of
perovskites.\cite{Bus99} On the other hand, it is possible that the more
unstable configurations of the octahedra correspond to the frozen tilt
waves. As a possible mechanism for peak A, we suggests the propagation of
kinks formed on the tilt waves. The formation and propagation of kink pairs
could be the elementary steps for bending a straight tilt wave or wall,
analogously to the case of dislocations.\cite{RF92} The formation of a kink
pair corresponds to the shift of a wall segment by a lattice unit
perpendicularly to the wall itself; it would require a relatively high
energy and could contribute to the slower relaxation peak T. Once formed,
the kinks could easily propagate or oscillate along the wall, involving the
switching of only few O octahedra or atoms.

Another simple mechanism which could explain the low temperature relaxation
is the tunneling of single O atoms. Indeed, EXAFS experiments\cite{HSH97}
suggest that the apical O atoms close to Sr dopants feel a double-well
potential in the $c$ direction. Peak A, however, is rather explainable in
terms of tunneling between minima that correspond to tilting of the
octahedra. In fact, it is present also in Sr-free samples and its
suppression by interstitial O and accelerated dynamics upon doping are
easily interpreted in terms of tilt modes.

Let us now consider the issue whether peak A is describable in terms of Eq.
(1) with appropriate expressions and distributions for the relaxation
strength $\Delta $ and rate $\tau $; this is usually assumed in many cases
of atomic tunneling, like light or off-centre atoms in crystals or the
two-level systems (TLS) in glasses.\cite{Esq98} A check which does not
require the knowledge of the forms of $\Delta $ and $\tau $ is to verify
that the high temperature side of the absorption peak divided by the
measurement frequency $f=\omega /2\pi $ is independent of $f$. In fact, if
Eq. (1) is true for each independent relaxation unit, with $\tau $ a
decreasing function of temperature, then at temperatures above the maximum
relaxation $\omega \tau $ becomes smaller than unity and can be discarded in
the denominator; therefore, for the high temperature side of the peak one
has $Q^{-1}/\omega \simeq \Delta \left( T\right) \,\tau \left( T\right) $,
independent of frequency. This is true for any form of $\Delta \left(
T\right) $ and $\tau \left( T\right) $ and even if integrated over
distributions of $\tau $ and $\Delta $, until the condition $\omega \tau \ll
1$ is satisfied for all the elementary relaxations. Figure 5b reports the
function $J=Q^{-1}T/f$ (which is proportional to the spectral density of
strain and therefore of the atomic motions, see later){\em \ }for peak A\ at 
$x=0.008$, after subtraction of a constant background shown in Fig. 5a. The
spectral densities measured at the three different frequencies merge on the
same curve at high temperature, indicating that Eq. (1), integrated over
distributions of $\tau $ and $\Delta $, is indeed appropriate for describing
peak A, at least for $x=0.008$.

The $J$ obtained for $x=0$ does not satisfy the same condition, because it
is described by a broad distribution of $\tau \left( T\right) $ which vary
with temperature at a slower rate than in the case $x>0$, so that the
condition $\omega \tau \ll 1$ is not satisfied in the high temperature side
of the peak for all the elementary relaxations. Therefore, from the
temperature position and shape of peak A we deduce that doping changes the
mean relaxation rate $\tau ^{-1}$ from a slowly varying function of
temperature in the undoped case to a function which increases faster with
temperature. The high temperature side of the spectral density in Fig. 4b
indicates that, for $x=0.008$, the rate $\tau ^{-1}$ increases faster than $%
T^{4}$ above 6$~$K.

The function $J$\ is proportional to the spectral density of the strain $%
\varepsilon $, namely the Fourier transform of the correlation function $%
\left\langle \varepsilon \left( t\right) \varepsilon \left( 0\right)
\right\rangle $, which in turn is directly related to the correlation
function of the displacements of the O atoms, or the tilts of the octahedra.
In fact, the absorption may be written as $Q^{-1}=S^{\prime \prime }/S,$
where in the present case the compliance $S$\ is the reciprocal of the
Young's modulus, and thanks to the fluctuation-dissipation theorem\cite{LL5}
on has 
\begin{equation}
S^{\prime \prime }=(\omega V\,/\,2k_{\text{B}}T)\,\int dt\,e^{-i\omega
t}\,\left\langle \varepsilon \left( t\right) \varepsilon \left( 0\right)
\right\rangle \,,  \label{S''}
\end{equation}
where $V$\ is the sample volume,\cite{WK96} or 
\begin{equation}
J=\frac{T}{\omega }Q^{-1}\propto \int dt\,e^{i\omega t}\left\langle
\varepsilon \left( t\right) \varepsilon \left( 0\right) \right\rangle \,.
\label{J}
\end{equation}
The fact that Eq. (1)\ describes peak A implies $\left\langle \varepsilon
\left( t\right) \varepsilon \left( 0\right) \right\rangle =\left\langle
\varepsilon _{0}^{2}\right\rangle \,e^{-t/\tau }$, which is due to atomic
displacements changing at an average rate $\tau ^{-1}$. In the tunneling
model, such displacements are associated with the transitions of the tunnel
system (TS) between its eigenstates. Such transitions are promoted by the
interaction between the TS and the various excitations of the solid,
generally consisting in emission and absorption of phonons and scattering of
the conduction electrons.\cite{Esq98} Instead, the resonant motion of the
atoms of the TS between the potential minima while remaining in the same
eigenstate produce a peak in the spectral density centred at the tunneling
frequency, which is different from Eq. (1) and would be observable only at
much higher frequencies.

An important difference between the relaxation process A and those due to
the TLS in glasses is that the latter are characterized by an extremely
broad distribution of parameters, mainly the tunneling energy $t$ and the
asymmetry between the minima of the double-well potential; instead, in La$%
_{2-x}$Sr$_{x}$CuO$_{4}$ the geometry of the tunnel systems is much better
defined, being some particularly unstable configurations of the octahedra.
As a consequence, the TLS relaxation in glasses produces a plateau in the
acoustic absorption and a linear term in the specific heat as a function of
temperature, whereas in La$_{2-x}$Sr$_{x}$CuO$_{4}$ we observe a well
defined peak in the absorption, and no linear contribution to the specific
heat.

\subsection{Interaction between the tilts of the octahedra and the hole
excitations}

We now argue that the marked acceleration of the local fluctuations with
doping (narrowing and shift to lower temperature of peak A) is the
manifestation of a direct coupling between the tilts of the octahedra and
the holes, similarly to the TS in metals, whose dynamics is dominated by the
interaction with the conduction electrons.\cite{Esq98} The transition rate
of a TS is generally of the form 
\begin{eqnarray}
\tau ^{-1}\left( T\right) &\propto &t^{2}f\left( T\right) \\
f\left( T\right) &=&f_{\text{ph}}\left( T\right) +f_{\text{el}}\left(
T\right) \,,
\end{eqnarray}
where $t$ is the tunneling matrix element and $f_{\text{ph}}\left( T\right) $
and $f_{\text{el}}\left( T\right) $ contain the interaction between the TS
and the phonons and electrons, whose excitation spectra depend on
temperature and hole density. Doping is expected to reduce the potential
barriers between the different tilts of the octahedra,\cite{61} and
therefore to increase $t$. This is explained in terms of a reduction of the
lattice mismatch between the CuO$_{2}$ planes and the La/Sr-O layers,\cite
{ZCG} which is thought to be the driving force for the octahedra tilting,
and the more evident manifestation is the decrease with doping of the
HTT/LTO transition temperature and the average tilt angle in the LTO phase.%
\cite{CCF,BS2} Therefore, the introduction of few percent of holes may
increase the tunneling frequency $t$, and possibly slightly modifies the
phonon spectrum, affecting also $f_{\text{ph}}\left( T\right) $; however,
these are expected to be minor changes, certainly not enough to cause the
qualitative change of the relaxation rate from the undoped to the $x=0.008$
case. The greater part of the enhancement of the relaxation rate has to be
attributed to the increased interaction with the doped holes, corresponding
to the term $f_{\text{el}}\left( T\right) $. Similar effects are observed on
the TS's in metals, whose dynamics is dominated by the interaction with the
conduction electrons.\cite{Esq98} For example, the relaxation rate of
interstitial H tunneling near an O atom in Nb sharply drops below the
critical tempeature,\cite{DG,MMW} when the opening of the superconducting
gap reduces the possibility of scattering of the electrons from the
tunneling particle. Models for the interaction between tunneling systems and
electrons have been developed and succesfully adopted, but the electronic
excitation spectrum of the cuprate superconductors is certainly different
from that of metals, and new models of the TS-electron interaction are
needed. Indeed, the asymptotic behaviour of $J\left( \omega ,T\right) $ at
high and low temperature for $x=0.008$ indicate that $f_{\text{el}}\left(
T\right) \sim T^{n}$ with $n\sim 3-5$; this differs from the temperature
dependence of $f_{\text{el}}\left( T\right) $ for tunnel systems in standard
metals, which is less than linear.\cite{Esq98}

It is even possible that the pseudo-diffusive lattice modes modify the
electron-phonon coupling. In fact, the characteristic frequencies of these
modes are far too slow for having any influence on the electron dynamics for 
$x\le 0.02$, but increase dramatically with doping. In terms of the
multiwell lattice potential, it is possible that at higher doping the
barriers between the minima become small enough to give rise to an
enhancement of the electron-phonon coupling predicted by some models with
anharmonic potentials.\cite{BBB} Such double-well potentials have been
searched for a long time, especially for the apical O atoms in YBa$_{2}$Cu$%
_{3}$O$_{6+x}$ (Ref. \onlinecite{MCB92}) and La$_{2-x}$Sr$_{x}$CuO$%
_{4+\delta }$ (Ref. \onlinecite{HSH97}) and are thought to play an important
role in determining the dominant mechanism of electron-phonon interaction.%
\cite{BBB}

\subsection{Static and dynamic tilt disorder}

The observation of an increased intensity of peak A at $x>0$ with respect to 
$x=0$ (Fig. 3) is important, since it provides an explanation for the
apparent discrepancy between the previous anelastic experiments on La$_{2}$%
CuO$_{4+\delta }$\cite{61} on one side and EXAFS\cite{HSH96} and atomic PDF%
\cite{BBK98} measurements on La$_{2-x}$Sr$_{x}$CuO$_{4}$ on the other side.
The acoustic technique sees a dynamic tilt disorder whose effect increases
with reducing $\delta $ (and therefore doping), while the latter techniques
see the opposite effect of an increasing tilt disorder with increasing $x$.
Actually, the EXAFS spectra and atomic PDF are sensitive to both static and
dynamic disorder; at $x=0$ the instantaneous fraction of octahedra swept by
the tilt waves is relatively small (few percents according to a crude
estimate\cite{68} of the NQR relaxation intensity of peak T) and its effect
on the PDF or EXAFS spectra is undetectable. At higher $x$, the
instantaneous tilt disorder will increase due to both the disorder in the
La/Sr sublattice and the lattice fluctuations. Instead, the anelastic
spectroscopy is sensitive only to the dynamic disorder with characteristic
frequency comparable to the measurement frequenct; the introduction of
interstitial O certainly increases the static tilt disorder but also
inhibits the dynamic one, resulting in the depression of peaks A and T
observed in Ref. \onlinecite{61}. By introducing substitutional Sr, which
disturbs the lattice much less than interstitial O, it is possible to
observe that doping actually increases the fraction of fluctuating
octahedra, and not only the static disorder.

\section{Conclusions}

It has been shown that in La$_{2-x}$Sr$_{x}$CuO$_{4}$ there are
pseudodiffusive tilt modes of the O octahedra which give rise to tunneling
systems. Such tunneling modes are already present in the undoped state but
their relaxation rate and intensity strongly increase with doping. The
enhancement of the rate of the local fluctuations is ascribable to the
direct coupling between the tilts of the octahedra and the electronic
excitations. The form of this interaction is different from the known cases
of coupling between tunneling systems and the conduction electrons.

\section*{Acknowledgments}

The authors thank Prof. A Rigamonti for useful discussions. This work has
been done in the framework of the Advanced Research Project SPIS of INFM.


\section{Captions}

Fig. 1 Normalized variation of the Young's modulus of four samples with
nominal $x=0$, 0.0075, 0.015 and 0.030 at the HTT/LTO transition. The
normalization is both in the amplitude and temperature position of the step.
The samples are labeled with the values of $x$ obtained from the combined
analysis of the steps in the modulus and in the absorption.

Fig. 2 Anelastic spectrum of La$_{2-x}$Sr$_{x}$CuO$_{4}$ ($x=0.019$) after O
reduction in vacuum, measured exciting three flexural modes. \label{Fig1}

Fig. 3 Anelastic spectra of reduced La$_{2-x}$Sr$_{x}$CuO$_{4}$ with $x=0$
(6.2~kHz), $x=0.019$ (7.2~kHz) and $x=0.032$ (9.2~kHz). \label{Fig2}

Fig. 4 Effect of the interstitial O present in the as-prepared state on
processes A and B. All the data refer to the 5th vibration modes (6.2~kHz
for $x=0$, 18~kHz for $x=0.019$, 22~kHz for $x=0.032$). The same symbols of
Fig. 2, are used for the reduced state, while the thick lines are for the
as-prepared state containing excess O.

Fig. 5 Peak A for $x=0.008$ measured at three vibration frequencies. The
dashed line in (a) is the background which was subtracted in order to derive
the spectral density $J=Q^{-1}T/\omega $ of process A in (b).

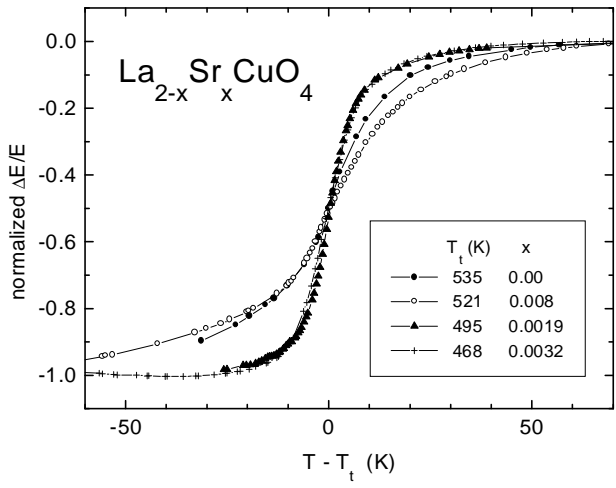
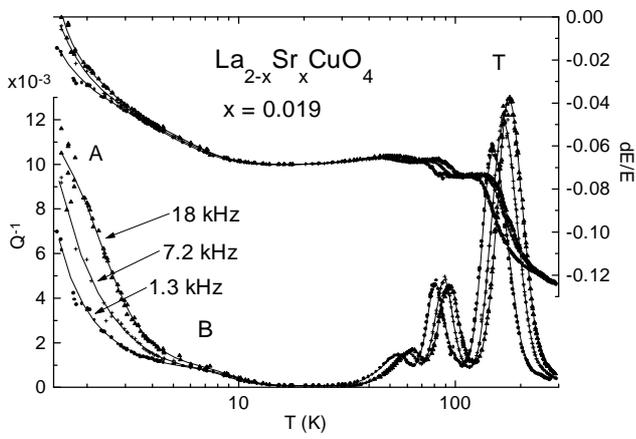
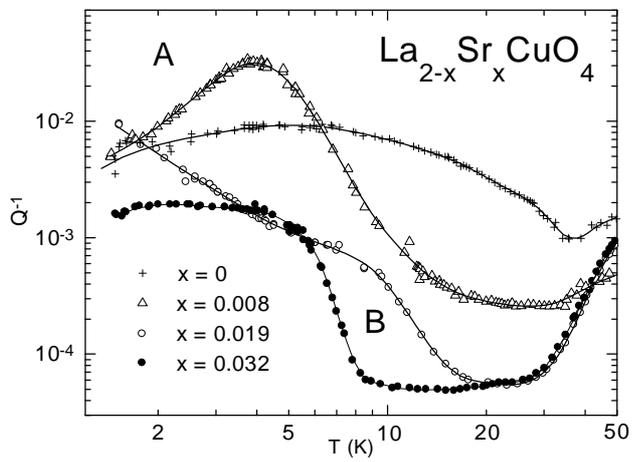
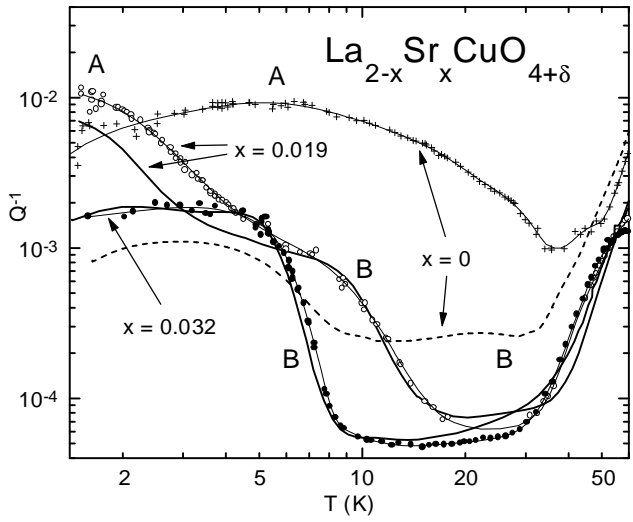
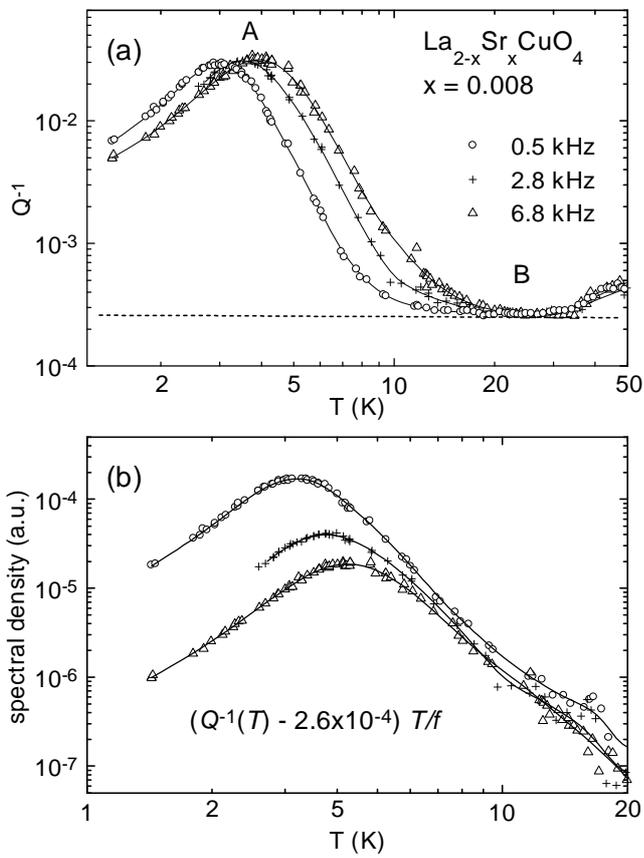
\end{document}